\documentstyle[psfig,conf_iap]{article}
\begin{document}

\newcommand{\fun}{\hbox{\ erg cm$^{-2}$ s$^{-1}$} }
\newcommand{\lun}{\hbox{\ erg s$^{-1}$} }
\newcommand{\lstar}{L^{\displaystyle\ast}_X}
\def\lesssim{\mathrel{\hbox{\rlap{\hbox{\lower4pt\hbox{$\sim$}}}\hbox{$<$}}}}
\def\gtrsim{\mathrel{\hbox{\rlap{\hbox{\lower4pt\hbox{$\sim$}}}\hbox{$>$}}}}
\null\vspace*{-3cm}\noindent \footnotesize{{\sf Proceedings of 
{\sl ``Wide Field Surveys in Cosmology"}, IAP Meeting, Paris, May 1998}} 
\normalsize\vspace*{2cm}
\pagestyle{plain}

\heading{%
%
X-ray Surveys of Distant Galaxy Clusters

%
} 
\par\medskip\noindent
\author{%
Piero Rosati $^{1}$
}
\address{%
European Southern Observatory, Garching b. M\"unchen, 85748 Germany
}

\begin{abstract}
I review  recent observational progress  in the search for and
study of distant galaxy clusters in the X-ray band, with particular
emphasis on the evolution of the abundance of X-ray clusters out to
$z\sim\! 1$.   Several on-going deep X-ray surveys
have led to the discovery of a sizeable population of clusters at
$z>0.5$ and have the sensitivity  to detect clusters beyond redshift
one.  These surveys have significantly improved our understanding of
cluster evolution  by showing that the
bulk of the population of galaxy clusters is not evolving significantly
since at least $z\simeq 0.8$, with some evolution limited to only the
most luminous, presumably most massive systems.  Thus far, a well
defined sample of very high redshift ($z\gtrsim 1$) clusters has been
difficult to assemble and represents one of the most challenging
observational tasks for the years to come.
\end{abstract}

\section{Introduction}
The redshift evolution of the
abundance of galaxy clusters has long served as a valuable tool with
which to test models of structure formation and set constraints on
fundamental cosmological parameters (e.g. \cite{eke96}, \cite{bah97}).
Being recognizaeble out to large redshifts, clusters are also ideal
laboratories to study the evolutionary history of old stellar systems,
such as E/S0s, back to early cosmic look-back times (e.g. \cite{sed97},
\cite{vdok98}).
It is therefore not surprising that a considerable observational effort
has been devoted over the last decade to the construction of
homogeneous samples of clusters over a large redshift baseline.  Until
a few years ago, however, the difficulty of finding high redshift
clusters in deep optical images and the limited sensitivity of early
X-ray surveys had resulted in only a handful of spectroscopically
confirmed clusters at $z>0.5$. As a result, the evolution of the space
densities of clusters, even at moderate look-back times, has been the
subject of a long-standing debate (e.g. \cite{cou91}, \cite{hen92}).

\section{Searches for X-ray clusters}

With the advent of X-ray imaging in the 80's, it was soon recognized
that X-ray searches for galaxy clusters have the advantage of revealing
physically-bound systems out to cosmologically interesting redshifts
and thus offer the unique opportunity to construct flux-limited samples
with well-understood selection functions.
Pioneering work in this field was carried out by Gioia et al.
(1990) and Henry et al. (1992) based on the  Einstein Medium
Sensitivity Survey (EMSS). By extending significantly 
the redshift range probed by previous
samples \cite{edg90} (based on non-imaging X-ray data), the EMSS survey 
has been for years the basis for several intensive follow-up studies (e.g.
CNOC survey \cite{Car97}).

 The ROSAT-PSPC detector, with its unprecedented sensitivity and
spatial resolution, made clusters
 high contrast, extended objects in the X-ray sky and has thus allowed for a
significant leap forward. ROSAT data have provided the means to carry
out large contiguous area surveys of nearby clusters with the ROSAT
All-Sky Survey (RASS) (\cite{ebe97}; B\"ohringer, this volume), as well
as much deeper serendipitous searches based on single pointings.
On-going X-ray surveys of distant galaxy clusters which utilize PSPC
archival data include the ROSAT Deep Cluster Survey (RDCS
\cite{ros95}, \cite{ros98}), the Serendipitous High-Redshift Archival
Rosat Cluster survey (SHARC \cite{col97}, \cite{bur97}), the Wide Angle
Rosat Pointed X-ray Survey of clusters (WARPS
\cite{sch97}, \cite{jon97}), the CfA large area survey (\cite{vik98},
\cite{vik98b}), and the RIXOS survey (\cite{cas95}). 
An additional survey is being carried out in the North
Ecliptic Pole (NEP, \cite{hen97}; Gioia, this volume), using the
deepest area scanned by the RASS.

\subsection{Strategies and Selection Functions}

Most  studies have adopted a similar methodology but somewhat
different strategies. Cluster candidates are selected from a
serendipitous search for {\sl extended X-ray sources} above a given 
flux limit in deep ROSAT-PSPC
pointed observations. Particular emphasis is given in these searches to
detection algorithms which are designed to probe a broad
range of cluster parameters (X-ray flux, surface brightness,
morphology) and to deal with the confusion effect at faint 
flux levels.  
A popular and well-suited approach is that of multi-scale analysis
based on wavelet techniques (e.g. \cite{ros95},\cite{vik98}).

\begin{figure}
\centerline{\vbox{
\psfig{figure=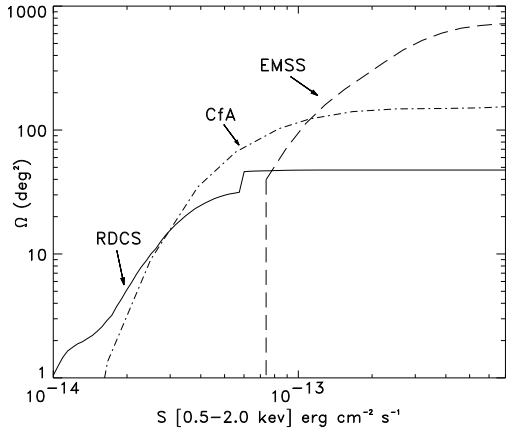,height=6cm}
}}
\caption[]{Comparison between the effective sky solid angle covered by
three cluster surveys as a function of the X-ray flux (EMSS
\cite{hen92}, CfA Survey \cite{vik98}, RDCS \cite{ros98}).
\label{fig:skycov}
}
\end{figure}

By covering different solid angles at varying fluxes these surveys
probe different regions in the X-ray luminosity--redshift plane (i.e. the
$N(L_X,z)$ distribution peaks at slightly different positions).
Fig.\ref{fig:skycov} illustrates the effective sky coverage of
the EMSS, compared to that of two ROSAT surveys (\cite{ros98},
\cite{vik98}). The EMSS has the greatest sensitivity to the most luminous,
yet most rare, systems but only a few clusters at high redshift lie above its
bright flux limit. On the other hand, deep ROSAT surveys  probe instead the
intermediate-to-faint end of the X-ray Luminosity Function (XLF).
As a result, they 
have lead to the discovery of many new clusters at $z\gtrsim 0.4$. The
RDCS, has pushed this search to the faintest fluxes yet, providing
sensitivity to the highest redshift systems (including $z\gtrsim 1$) 
with $L_X\approx  \lstar$,
whereas the CfA survey has covered a significantly larger area at high
fluxes thus probing the interesting bright end of the XLF at $z\lesssim 0.6$.

Extensive optical follow-up programs associated with these surveys have,
to date, lead to the identification of roughly 200 new clusters or groups,
and have increased the number of clusters known at $z>0.5$ by about
a factor of five.  As an
example, out of more than 100 clusters spectroscopically identified in
the RDCS, roughly one-third lie at $z>0.4$ and a quarter at $z>0.5$.  The
fact that very few have been discovered so far at $z>0.85$ is not due
to a lack of sensitivity of X-ray searches at these redshifts, but
rather reflects the difficulty of carrying out the spectroscopic
confirmation with 4m-class telescope.

Since cluster candidates in such surveys are selected on the basis of
their spatial extent, a challenging task is to understand
 and quantify  selection effects at varying fluxes.  With the PSPC PSF
degrading rapidly at large off-axis angles across the detector, the
survey  becomes surface brightness limited below a given flux. This
important effect can be accounted for by modelling the sky coverage of
a given survey as function of flux and intrinsic size of the clusters
 (fig.\ref{fig:skycov}). An
overestimate of the solid angle covered at low fluxes and its
corresponding search volume can lead to overestimating the amount of
evolution of the cluster population (\cite{cas95}). Furthermore, the
surface brightness ($\Sigma$) dimming at high-$z$ can be a serious
source of incompleteness in the faintest flux bins and depends
critically on the unknown steepness of the $\Sigma$-profile of X-ray
clusters at high redshift, as well as its evolution.  Again, the
task of the observer is to understand the X-ray flux in a given
survey below which this effect becomes important.  An additional source
of incompleteness, which will be difficult to quantify until the next
generation of high resolution X-ray imagers become available, may be
caused by clusters hosting X-ray bright AGN.  A discussion of the
methods which are most effective in quantifying the selection function
of X-ray surveys goes beyond the purpose of this review.

On purele empirical grounds, the importance of these effects will
become apparent when it will be possible to compare the number
densities of distant clusters selected of the basis of their angular
extent with the NEP survey, which set out to identify {\sl all} the X-ray
sources down to a given flux over a 80 deg$^2$ area, regardless of
their spatial extent.

\section{Evolution of the Cluster Abundance out $z\simeq 0.8$}
 
One of the primary goal of the aforementioned X-ray surveys is to study the
redshift evolution of the cluster abundance at a given X-ray
luminosity.  This is characterized by the $z$--dependent XLF or its
projections along the redshift and flux axis, respectively, i.e. the number
counts, $N(>\! S)$, and the redshift distribution, $N(z)$.
Such distribution functions of observables can then be directly
compared with theories of structure formation. 

\subsection{The Local XLF}
The determination of the local ($z\lesssim 0.3$) XLF obviously plays a
crucial role in assessing the evolution of the cluster abundance at
higher redshifts and much progress has recently been made in this direction.

\begin{figure}[h]
\centerline{\vbox{
\psfig{figure=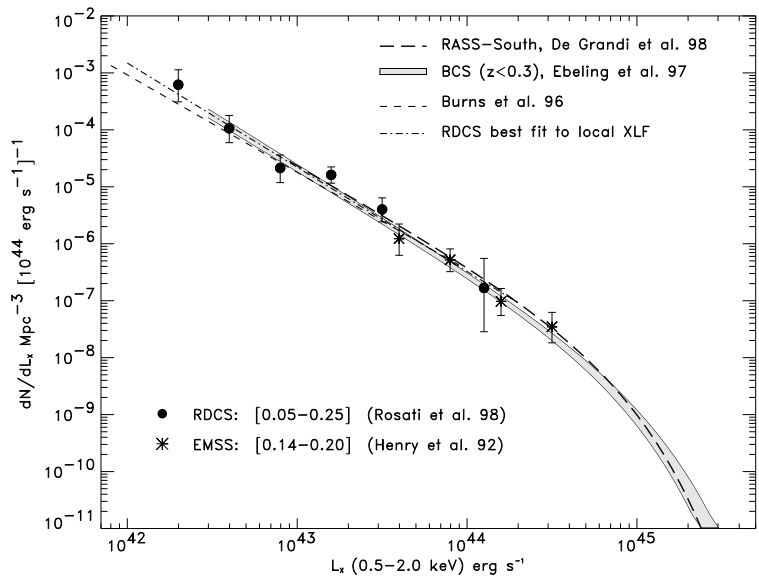,height=8cm,angle=90}
}}
\caption[]{Local X-ray Luminosity Function of clusters from different surveys.
\label{fig:locxlf}
}
\end{figure}

Fig.\ref{fig:locxlf} shows different determinations of the local cluster 
abundance.
The Brightest Cluster Survey (BCS) \cite{ebe97} covers a large $L_X$
range, similar to an independent RASS survey in the southern sky
(\cite{sdg98} and B\"ohringer, this volume). Complementing data are
provided by the 
RDCS and the survey by Burns et al. (\cite{B96}) which probes 
the very faint end.
An excellent  agreement is apparent between {\sl all} these independent
determinations, all having  faint end slopes in the range $1.75-1.85$ and
consistent normalizations. This is quite remarkable considering that
all these surveys used completely different selection techniques (from
pure optical to pure X-ray) and independent datasets. 
This situation contrasts with that exsisting only two years ago, when 
different surveys were finding faint
end slopes in the range $1.1-2.2$. This discrepancy was 
possibly due to the completeness levels and sky
coverages of  early samples which were not fully understood.
It would thus appear that the local cluster abundance, $N(L_X,z\simeq 0)$,
is now well established and can be safely used as a reference for
studying the evolution at higher redshifts.
Moreover, the BCS analysis at $z<0.3$ \cite{ebe97} shows that the 
evolution of the bright end found by the EMSS at $z> 0.3$ (fewer high
luminosity clusters) does not extend to lower redshifts.

\subsection{The Cluster Log$N$-Log$S$}

\begin{figure}
\centerline{\vbox{
\psfig{figure=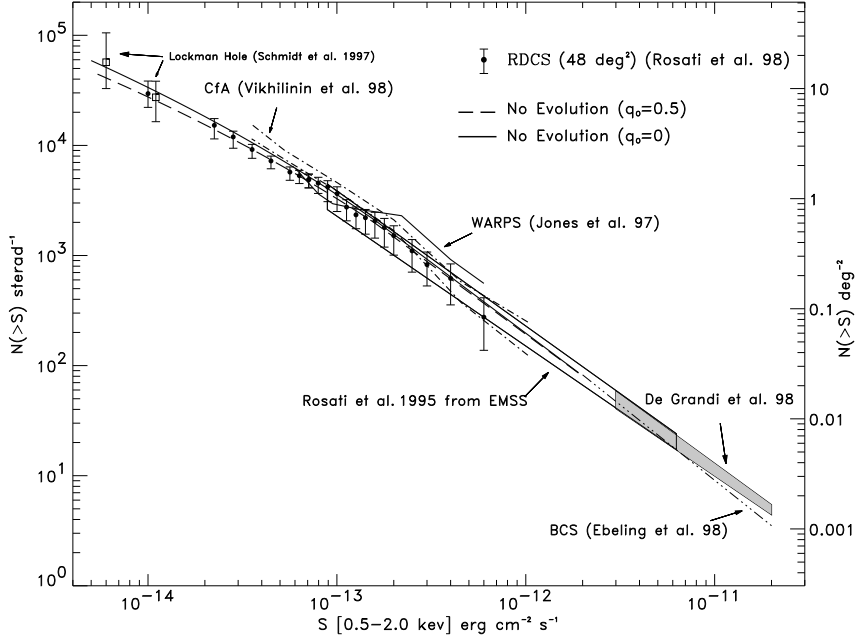,height=9cm,angle=90}
}}
\caption[]{The observed cluster cumulative number counts from various sources.
\label{fig:ngts}
}
\end{figure}

A summary of the observed cumulative cluster number counts  is given in
fig.\ref{fig:ngts}. This compilation includes both shallow and deep
surveys (CfA, RDCS, WARPS) so as to cover more than three decades in flux.
Once again,  we note an encouraging agreement at the $2\sigma$ level among
independent determinations. The slight difference between the RDCS and
the Vikhlinin et al. survey at low fluxes may be due to different
prescriptions used in these samples to evaluate the ``total flux" of the
clusters, a measurement which inherently depends on the assumed
$\Sigma$-profile of a cluster in background limited regime.
Although the Log$N$--Log$S$ is not a very robust diagnostic tool to
investigate the evolution of the cluster population, particularly for
the most luminous, rare systems, we note that the observed counts are
consistent with no-evolution predictions based on a fit of the local XLF
in fig.\ref{fig:locxlf}.

\begin{figure}
\centerline{\vbox{
\psfig{figure=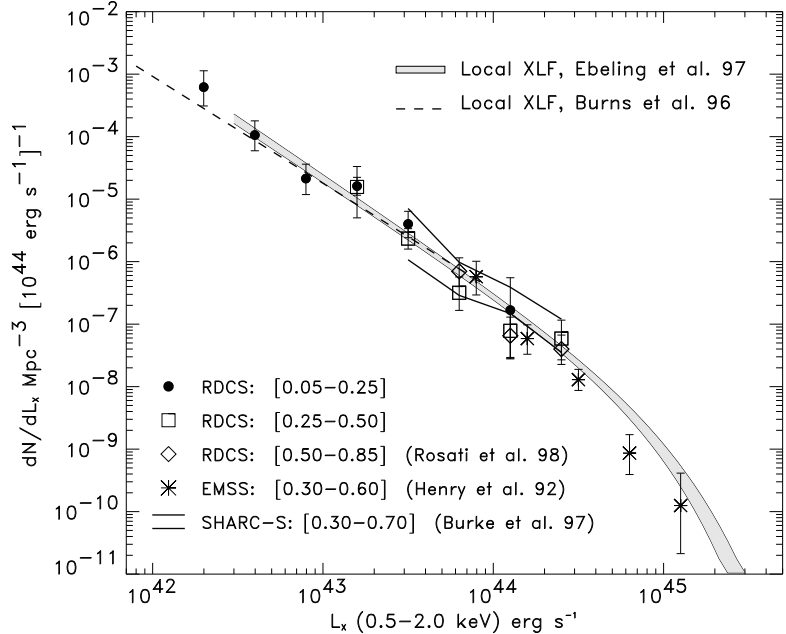,height=9cm}
}}
\caption[]{The X-ray Luminosity Function of distant clusters compiled from
 various sources and compared with local XLFs.
\label{fig:distxlf}
}
\end{figure}

\subsection{The Cluster XLF at higher redsfhits}

Moving to higher redshifts, several measurements of the XLF can now be
compared with the original determination of the XLF in the range $z=[0.3-0.6]$
from the EMSS. Fig.\ref{fig:distxlf} shows that the 
number densities from different samples
are in general in very good agreement in
regions of overlapping luminosities. Further inspection of the XLF in
bins of increasing redshift fails to show any significant evolution out
to $z\simeq 0.8$ (\cite{ros98}).
By combining independent analyses based either on the XLF
(\cite{bur97}, \cite{ros98}) or the Log$N$--Log$S$ (\cite{jon97}, \cite{ros98},
\cite{vik98b}), it emerges that the volume density of clusters per unit
luminosity has remained constant within the present uncertainties, over a
wide range in luminosities ($2\times 10^{42} \lesssim L_X(\lun)
\lesssim 3\times 10^{44}$). This $L_X$ range encompasses
 the bulk of the cluster
population, from poor groups to moderately rich clusters with
$L_X\simeq \lstar\approx 4\times 10^{44}\lun $ (roughly the Coma cluster).
These results are not in conflict with the EMSS findings of a
steepening of the XLF at luminosities {\sl in excess} of the local $\lstar$,
a result which is consistent with the more recent analysis of the CfA survey
in the highest luminosity bin \cite{vik98b}.

\begin{figure}
\centerline{\vbox{
\psfig{figure=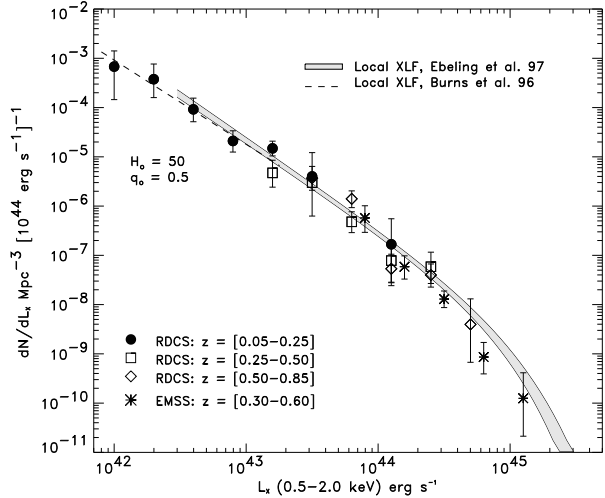,height=7cm}
}}
\caption[]{The latest determination of cluster X-ray Luminosity Function
 from the RDCS (81 clusters with $F_X>3.5\times 10^{-14}\fun$). 
\label{fig:newxlf}
}
\end{figure}

The latest cluster XLF derived from a flux-limited sample 
($F_X${\small [0.5-2.0 keV]}$ >3.5\times 10^{-14}\fun$) of 81 clusters spectroscopically
identified in the RCCS survey is shown fig.\ref{fig:newxlf}. The
picture described above is further confirmed. It also appears that in
order to make a significant step forward in understanding the evolution
of the most luminous systems at high redshifts ($z>0.7$), a new survey
covering at least 10 deg$^2$ {\it at} $F_X\simeq 1\times 10^{-14}\fun$
is needed. This is within reach with several years of serendipitous pointings
to be accumulated with XMM and AXAF.

\section{Cluster searches at $z\gtrsim 1$}
\begin{figure}
\centerline{\vbox{
\psfig{figure=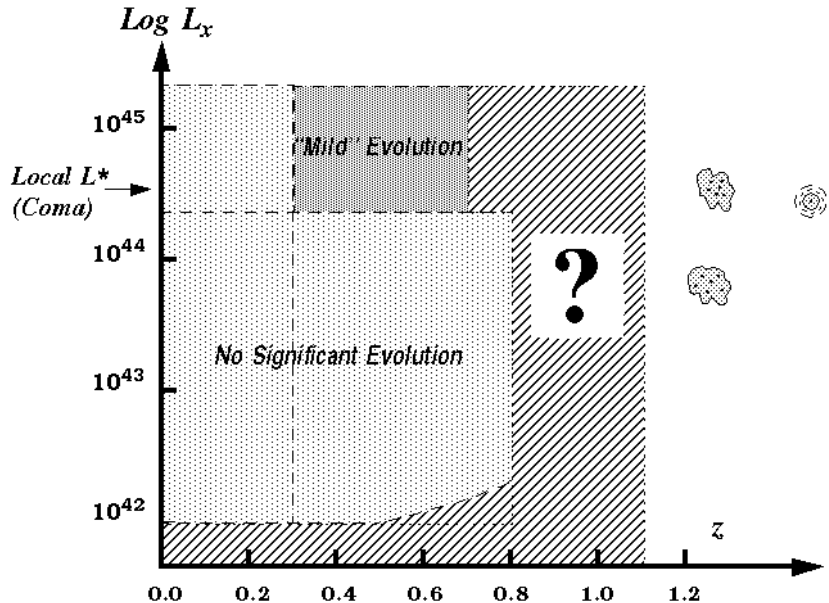,height=9cm}
}}
\caption[]{Cartoon summarizing the observational status of X-ray
cluster evolution. Very little is presently  known about the
cluster population at $z>0.8$, although bona-fide
X-ray clusters have been detected out to $z=1.27$ \protect\cite{sta97} 
\protect\cite{ros99} and
diffuse X-ray emission detected around radio sources to even higher
redshift \protect\cite{dic98}.
\label{fig:evol}
}
\end{figure}

Fig.\ref{fig:evol} summarizes our current understanding of
 cluster evolution in the ($L_X, z$) plane at the end of the
ROSAT era.  
A major unknown concerns the cluster abundance beyond redshift of one.
If one assumes that the evolutionary trend in the
cluster population continues past $z=1$, the observed $N(L_X,z)$
can be extrapolated (taking also into account the estimated
incompleteness at the faintest flux levels) to predict that ROSAT-PSPC
searches must be still sensitive to the feeble X-ray emission 
from $\lstar$ clusters at $z>1$. 
For example, in the RDCS one would expect up to a dozen X-ray
luminous clusters at $z\gtrsim 1$. However, in order to identify these
clusters, near-IR deep imaging and spectroscopy with
8m-class are required. 

The efficacy of near-IR searches  for
high-$z$ cluster searches has recently been proven by 
Stanford et al.~\cite{sta97} who have identified a cluster at $z=1.27$ in 
a near-IR field galaxy survey. 
A corresponding
extended X-ray source was found by the same authors in a deep ROSAT
pointing, as well as serendipitously in the RDCS candidate sample.
More recently, another RDCS faint candidate has been confirmed at
$z=1.26$ using IR imaging and Keck spectroscopy \cite{ros99}. The X-ray luminosities of both these systems are around $10^{44} \lun$ 
in the [0.5-2.0 keV] band.
These findings show that the combination of deep X-ray observations and
near-IR imaging is an efficient method by which to identify massive clusters 
at $z\gtrsim 1$ in a {\sl serendipitous fashion}, thus allowing
statistical estimates of the cluster abundance to be made. The much improved
sensitivities of XMM and AXAF will make this method particularly attractive.
 
At even higher redshifts, a viable method to identify clusters is to
target powerful radio sources (e.g. \cite{dic98}). Deep ROSAT pointings
on these sources have revealed the existence of diffuse X-ray mission
out to $z\simeq 1.8$ which most likely arises from
hot intra-cluster gas trapped in deep cluster potential wells at such early epochs \cite{dic98}.

\section{Discussion}

Remarkable observational progress has been made over the last years 
in determining the
abundance of galaxy clusters out to $z\sim\! 1$, as is
underscored by the convergence of the results from several independent
studies. At the begining of the ROSAT era, only a few years ago,
controversy sorrounded the usefulness of X-ray
surveys of distant galaxy clusters. This prejudice arose from an
overinterpretation of the early results of the EMSS survey which, as of
today, remain basically correct. Although in the early analysis of
Gioia et al. \cite{gio90} it was clearly stated that the evolution of
the XLF was limited {\sl only} to the very luminous systems, this detail
was often overlooked for many following years.
Indeed, this evolution was believed to
extend through the bulk of the cluster population ($L_X\lesssim\lstar$)
not adequately probed by the EMSS at high redshifts.
The original contoversy concerning cluster evolution 
inferred from optical and X-ray data finds a possible explanation in
this. 
Optical surveys (\cite{cou91},\cite{pos96}) have shown no dramatic
decline in the coomoving volume density of rich clusters out to
$z\simeq 0.5$. This was considered to be in contrast with the EMSS
findings. However, these optical searches covered limited solid angles
(much smaller than the EMSS)  and therefore did not probe adequately
the seemingly evolving high end of the cluster mass function.

\begin{figure}
\centerline{\vbox{
\psfig{figure=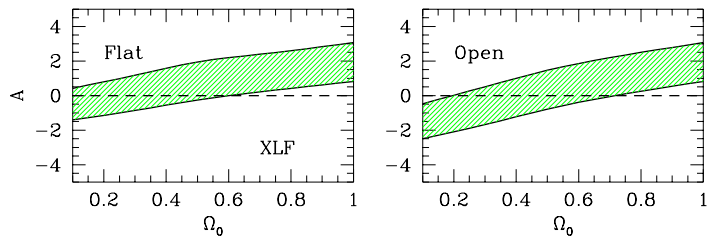,height=10cm}
}}
\vskip -6.5cm
\caption[]{Constraints at 90\% c.l. on the $\Omega_0-A$ plane by
matching the XLF as a function of redshifts from various X-ray surveys
\protect\cite{bor98}. The parameter $A$ describes the redshift
evolution of the X-Luminosity-Temperature relation: $L_X\sim (1+z)^A T^\alpha$. 
\label{fig:omega}
}
\end{figure}

The theoretical interpretation of the new results on the evolution of the
cluster abundance is still ambiguous.  The implications that these
findings have for models of cluster
formation have been discussed by several authors (e.g. \cite{ks97},
\cite{me98}, \cite{cav98}, \cite{bor98}, \cite{bow98}). By following a
phenomenological approach,  one can constrain cosmological
parameters and evolutionary parameters of the intra-cluster medium by
matching models with observed distributions, such as
$N(L_X,z)$, $N(z)$, $N(>\!S)$.  
This analysis has shown \cite{bor98} that without additional
observational inputs from the temperatures of high-$z$ clusters or a
better understanding of the physics governing the evolution of their
gaseous component, it is difficult to draw firm conclusions on the
value of the density parameter $\Omega_0$. 
As an example, fig.\ref{fig:omega} shows
the degeneracy between $\Omega_0$ and the evolutionary parameter of the
X-Luminosity-Temperature (L--T) relation. Recent measurements of 
cluster temperatures at moderate redshifts indicate that the L-T relation
does not evolve significantly
out to $z\simeq 0.5$ (\cite{SM97},\cite{hen97}) (i.e. $A\approx 0$), which
would favour a low-$\Omega$ universe.

\section{Future Prospetcs}

The next decade promises to be particulalry exciting for cluster astrophysics.
The new ROSAT samples described herein will figure prominently in studies
for years  to come.
The new generation of optical and near-IR mosaic imagers and highly
efficient multiplexing spectrographs on 8-meter class telescopes (Keck,
VLT, GEMINI) will probe a region of the parameter space of
redshift, solid angle and limiting flux which has been 
completely unexplored with 4m-class telescopes and conventional detectors.
The Advanced Camera aboard HST (2000) will add sub-kpc morphological 
information to this multi dimensional data set and will permit detailed
studies of their lensing patterns.  
The impact of AXAF, XMM, and Sunyaev-Zeldovich measurements are described 
elsewhere in this volume (see contributions from K.Romer; M.Pierre; J.Bartlet).

The next generation of $X$-ray satellites and the already available
large optical telescopes should open the possibility of determining
masses of distant clusters via $X$-ray temperature measurements, virial
analysis and gravitational lensing studies. Carrying on such
observations for an even limited number of clusters, extracted from a
well defined statistical sample, will determine both the evolution of
the cluster internal dynamics and the value of the cosmological density
parameter.

\acknowledgements{I am particularly grateful to Roberto Della Ceca,  
Colin Norman, Riccardo Giacconi and Stefano Borgani, 
who contributed to part of the work presented here.
}

\vfill
\end{document}